\documentclass[preprint,showpacs,aps,prb,psfig]{revtex4}
\usepackage{graphicx}

\begin{document}
\title{Fast noise in the Landau-Zener theory.}
\author{V.L. Pokrovsky$^{1,2}$, N.A. Sinitsyn$^1$}
\address{$^1$Department of Physics, Texas A\&M University, College Station,\\
Texas 77843-4242,\\
$^2$Landau Institute of Theoretical Physics, Chernogolovka, Moscow region\\
142432, Russia}
\date{today}

\begin{abstract}
We study the influence of a fast noise on Landau-Zener transitions.
We demonstrate that a fast colored noise much
weaker than the conventional white noise can produce transitions
itself or can change substantially the Landau-Zener transition
probabilities. In the limit of fast colored or strong white noise
we derive asymptotically exact formulae for transition
probabilities and study the time evolution of a spin coupled to
the noise and a sweeping magnetic field.
\end{abstract}
 \maketitle


\section{Introduction}

Landau-Zener (LZ) formula for transition probabilities at avoided
crossing of two levels is one of a few fundamental results of
non-stationary quantum mechanics. Its rather general character and
simplicity makes it extremely suitable for versatile applications.
Traditionally it was applied in quantum chemistry \cite{chemistry}
and in collision theory \cite{inc}, \cite{nik}. A recent treatment
of the experiments on the quantum molecular hysteresis in
nanomagnets by Wernsdorfer and Sessoli \cite{{WS1},{WS2},{nm}} was
a real triumph of the LZ theory. A substantial contributions to
the theory of spin tunnelling in these molecules was made by
theorists \cite{{Garg},{GC},{Prokofev}}. Landau-Zener formula and
its generalizations were recently employed also for charge
transport in nanostructures
\cite{{av1},{av2},{Gef},{Iliescu},{dot2}}, Bose-Einstein
condensates \cite{Y1} and quantum computing \cite{FG}.

Extensions of the LZ theory to the case of multilevel crossing are
less general. Nevertheless, some of them were realistic enough to
justify remarkable efforts on the side of theorists for their
analysis. Level correlations and localization in energy space were
studied in \cite{Gef-loc}. The pioneering work by Demkov and Osherov \cite{demkov}
treated exactly the crossing of a single level with a band of parallel
levels. In the work \cite{zeeman2} Hioe and Carrol solved a
problem of transitions in a Zeeman multiplet of an arbitrary spin
S in a magnetic field with a constant perpendicular component and
a time-dependent parallel component passing through zero value.
Numerous generalizations of these results were found
\cite{{brand},{deminf1},{bow},{dem33},{zeeman1},{dem3},{PS},{usuki}}.
A general point of view on all these exactly solvable models
proposed by one of authors \cite{sinitsyn} allowed to construct an
algorithm for series of new solvable models. Another extensions
include nonlinear LZ-model \cite{{Liu},{Liu2},{Zob}} and LZ problem with
nonlinear sweep \cite{Garanin}. To apply the LZ formula and its
multi-state extensions to real systems it is often necessary to
take into account the interaction with environment. Such attempts
were made in a series of works
\cite{{GC},{Sai},{kay4},{Kay},{kay3},{Nish},{SaiKay},{Ao},{Kob},{Shi},{Kay2},{Loss},{Ben},{Usuki2}}, however the problem was not
solved completely. Kayanuma {\it et al.} \cite{{Sai},{kay4},{Kay}}
have obtained an elegant analytic result for the diagonal white
noise. The non-diagonal colored noise was considered by Kayanuma
\cite{kay3} for the two-level crossing without a constant coupling
term. He has found the transition probability in the limit of
infinitely short noise correlation time. His result was disputed
by Nishino {\it et al.} \cite{Nish}. On the basis of their
numerical calculations these authors concluded that the
transitions due to non-diagonal noise are weak in many realistic situations. The
discrepancy is associated with different limiting
procedures in calculations. In the limit of infinitely short correlations 
but finite amplitude of sweeping field the transition probability was found to be vanishingly small.

 The influence of the noise onto the
multilevel crossing was studied so far in only one work by Saito
and Kayanuma \cite{SaiKay}, who considered the 3-level crossing at
a special relations between parameters in the limit of strong
decoherence.

  The purpose of this article is to present a systematic study
of the influence of noise, including the colored noise, onto the
LZ transitions and to generalize it to multistate LZ problems. We
demonstrate that the  LZ transitions are sensitive to the colored
noise much weaker than usual $\delta$-like white noise. The latter
can be considered as a limit of a noise whose correlation time
goes to zero and simultaneously its square of amplitude goes to
infinity, so that their product remains a constant. We prove that
such a white non-diagonal noise always leads to equal population
of the crossing levels. However, the noise, whose correlation time
goes to zero, but its amplitude remains a constant, produces
non-trivial transition probabilities as it was first found by
Kayanuma \cite{kay3} for a special type of the noise correlation
function. 
Another subtle problem is the order of limiting
processes \cite{Nish}. The resulting probabilities depend crucially on what
happens first: time asymptotically goes to infinity or correlation
time goes to zero. Analysis of these problems allowed us to
reconcile works \cite{kay3} and \cite{Nish}. In our work we first
find simple analytical result for a transition produced by a most
general short-time correlated noise in 2-level systems and the
change of the LZ probabilities produced by such a noise. We check
these analytical results by numerical calculations. We also study
the influence of the noise on transitions at multilevel crossing.

The plan of the article is as follows. In section II we generalize
the result of Kayanuma for transverse noise \cite{kay3} to the
case of the arbitrary Gaussian noise in all three directions. Next
we demonstrate its generalization to a three level system. In the
section IV we study the time dependence of the density matrix with
LZ transitions stimulated by fast noise. In the fifth section we
propose a formula that incorporates constant transverse magnetic
field and compare its predictions with numerical simulations. In
section VI we consider the master equation for an arbitrary spin
placed into a regular varying field and noisy magnetic field along
the $z$-direction and a constant field along $x$-direction and
find simple expressions for transition probabilities in the limit
of a strong decoherence. In section VII we perform similar
calculations for a charged particle on a periodic chain driven by
a time-dependent electric field and compare our results with those
for a completely coherent evolution.

\section{Colored noise in two level LZ transitions.}

LZ transitions in a two-level system with a  non-diagonal noise
were studied by Kayanuma in \cite{kay3}. The Hamiltonian of the
problem was chosen to be
\begin{equation}
  H=\beta t \sigma _z + \eta _x \sigma _x
  \label{hk}
\end{equation}
where $\eta _x$ is  the noise field with the correlation function
$\langle\eta _x(t_1) \eta _x (t_2)\rangle =J_x^2
e^{-\lambda_x|t_1-t_2|}$ and $\sigma_i$ are Pauli matrices. In the
limit of infinitely short correlation time $\lambda \rightarrow
\infty$ Kayanuma has found a simple analytical result for the
transition probability.

The choice of Kayanuma corresponds to the spin $1/2$ problem with
noisy magnetic field along the $x$-axes only. We generalize the
Kayanuma model introducing a more general Hamiltonian with all 3
components of random magnetic field being non-zero and with a most
general form of the short-time correlation tensor:
\begin{equation}\label{hi}
H=\beta t \sigma _z + \sum _i \eta _i (t) \sigma_i,\,\, i=x,y,z
\end{equation}
\begin{equation}
\langle\eta_i (t_1) \eta _j (t_2)\rangle =
g_{ij}(\lambda|t_1-t_2|) \label{correlator}
\end{equation}
We assume that though $g_{ij}$ can be different for different $i,j$,
they are of the same order of magnitude. We consider the limit of
fast noise with $\lambda \rightarrow \infty$, where $\lambda$ is
the inverse characteristic decay time of the correlator $g_{ij}$.

The density matrix elements for the system with the Hamiltonian
(\ref{hk}) obey the following system of ordinary differential
equations:

\begin{equation}\label{dm5}
  \begin{array}{l}
     i\dot{\rho}=2(-\rho_{12}(\eta_x - i \eta_y)+\rho_{21} (\eta_x +
    i \eta_y))\\
    i\dot{\rho}_{12}=2 (\beta t + \eta_z)\rho_{12} - \rho (\eta_x + i
    \eta_y)\\
    i\dot{\rho}_{21} = -2(\beta t+\eta_z) \rho _{21}+\rho
    (\eta_x-i\eta_y)
  \end{array}
\end{equation}
where $\rho =\rho_{11}-\rho_{22}$. By elimination of non-diagonal
matrix elements equations (\ref{dm5}) are transformed into one
integral-differential equation for the occupation difference
$\rho$:
\begin{equation}\label{ie1}
\begin{array}{l}
  \frac{d\rho(t)}{dt}=-2 e^{-i\beta t^2} (\eta_x (t) - i
  \eta_y (t)) \int _{-\infty}^{t} dt_1 e^{i \beta t_1^2-i\int_{t_1}^t\eta_z(t_2)dt_2}
  (\eta_x (t_1) + i\eta_y (t_1))\rho (t_1) \\
   -2 e^{i \beta t^2} (\eta_x (t) + i
  \eta_y (t)) \int _{-\infty}^{t} dt_1 e^{-i \beta t_1^2+i\int_{t_1}^t\eta_z(t_2)dt_2}
  (\eta_x (t_1)  - i\eta_y (t_1))\rho (t_1)
\end{array}
\end{equation}

The solution of this equation can be formally found as an infinite series
in powers of $\eta_i$ that must be averaged over noise
realizations. A typical term contains the product $\eta_{i_1}(t_1)
\eta _{i_2}(t_2)\ldots \eta _{i_n}(t_n)$. Its average is equal to
the sum of all possible products of pair correlators since we
assume the noise to be Gaussian. Kayanuma \cite{kay3} has shown
that in the limit of very fast noise only the term in which the pairing
is ideally ordered in time, i.e. the pairs are
(12)(34)...(2n-1,2n), contributes a finite value into the
integral. Other pairings contribute terms, which are by a power
of infinitely small parameter $1/\lambda$ smaller and can be
neglected. This Kayanuma's observation is completely analogues to
the theorem proven by Abrikosov and Gor'kov in their theory of
impurities in a metal \cite{AbrGor}. Note that the diagonal
component of noise is inessential in this approximation and can be
omitted. These facts allow to write down the integral-differential
equation for the average value of $\rho$ as follows:
\begin{equation}
\frac{d\langle\rho\rangle (t)}{dt}=-4\int_{-\infty}^t\cos\left(\beta(t^2-t_1^2)\right)
F\left(\lambda|t-t_1|\right)\langle\rho\rangle (t_1)dt_1
\label{int-diff}
\end{equation}
where $F=g_{xx}+g_{yy}$. Now we can employ the approximation of
the fast noise assuming that the average $\langle\rho\rangle (t)$
almost does not change in the interval of time  $1/\lambda$ and
that integral of correlation function is convergent. In this
approximation we can extract $\langle\rho\rangle (t)$ from the
integral in the right-hand side of equation (\ref{int-diff}) and
expand the argument of the cosine near the end point $t_1=t$ of
the integral. The resulting differential equation reads:
\begin{equation}
\frac{d\langle\rho\rangle}{dt}=-4\hat{F}(2\beta
t)\langle\rho\rangle  \label{diff}
\end{equation}
where $\hat{F}(q)$ is the cosine Fourier transform of the function
$F$:
\begin{equation}
\hat{F}(q)=\int_{-\infty}^{\infty}\cos(q\tau)F(\lambda|\tau|)d\tau
\label{Fourier}
\end{equation}
Note that the characteristic value of $q$ are of the order
$\lambda$ and, respectively, essential values of $t$ in equation
(\ref{diff}) are $t\sim\lambda/\beta$. We see that essential
values of time go to infinity together with $\lambda$. It shows
that the order of limiting processes is indeed very important.
Here we first calculate the transition probability, i.e. the
diagonal elements of the density matrix at $t=\infty$ and at very
large, but still finite $\lambda$ and put it infinity in the end.
Solving equation (\ref{diff}), we find:
\begin{equation}
\langle\rho\rangle (t) = \langle\rho\rangle
(-\infty)\exp\left(-4\int_{-\infty}^t\hat{F}(2\beta
t^{\prime})dt^{\prime}\right) \label{exp}
\end{equation}
Together with the standard equation $\rho_{11}+\rho_{22}=1$
equation (\ref{exp}) determines average occupation number of each
level at any time. The most interesting are the transition
probabilities at $t=+\infty$, which can be obtained from the same
equation (\ref{exp}). Note that the integral in the exponent at
$t=+\infty$ becomes equal to
\begin{equation}
\int_{-\infty}^{\infty}\hat{F}(2\beta t^{\prime})dt^{\prime}=
\frac{\pi}{2\beta}F(0) \label{infty}
\end{equation}
Thus, from equations (\ref{exp}) and (\ref{infty}) we find:
\begin{eqnarray}
\rho_{11}(+\infty)=\frac{1}{2}\left[1+\left(2\rho_{11}
(-\infty)-1\right)\exp\left(-\frac{2\pi
F(0)}{\beta}\right)\right];\label{11} \\
\rho_{22}(+\infty)=\frac{1}{2}\left[1-\left(2\rho_{11}
(-\infty)-1\right)\exp\left(-\frac{2\pi F(0)}{\beta}\right)\right]
\label{22}
\end{eqnarray}
Putting $\rho_{11}(-\infty)=1$, we find the transition
probability:
\begin{equation}
P_{1\rightarrow 2}=\rho_{22}(+\infty)\mid_{\rho_{11}(-\infty)=1}=
\frac{1}{2}\left[1-\exp\left(-\frac{2\pi
F(0)}{\beta}\right)\right] \label{probability}
\end{equation}
For the standard white noise the correlators
$\langle\eta_i(t)\eta_j(t^{\prime})\rangle$ turn into
delta-functions in the limit $\lambda=\infty$. It means that their
values at $t=t^{\prime}$ become infinitely large. In particular it
means that $F(0)$ is infinitely large for the standard white
noise. In this case, as it is seen from equations (\ref{11}),
(\ref{22}) the occupancies of both levels are equal to 1/2. Thus,
the standard white noise leads to complete loss of initial state
memory and equipartition of the levels. On the other hand, if the
amplitude of noise remains finite, the occupation numbers
conserve memory on the initial state.

In the limit of the fast noise the transition probabilities are
determined only by the average square of non-diagonal noise and
are not sensitive to the diagonal noise. To illustrate this
statement we consider a special case when the diagonal noise does
not correlate with the non-diagonal one. Then, as it is seen from
equation (\ref{ie1}), the averaging over the $z$-component of the
noise leads to multiplication of the coefficients in this
integral-differential equation by the Debye-Waller factor
$$
\langle\exp \left( \int_{t_1}^t\eta_z(\tau)d\tau \right) \rangle
$$
$$
= exp \left( -\frac{1}{2 \lambda } (t-t_1)
\int_{-\infty}^{\infty}
 g_{zz} (\theta) d \theta \right)
$$
An essential interval of integration over $t_1$ near $t_1=t$ is
$1/\lambda$. In this interval the Debye-Waller factor changes by
the value $\sim 1/(\lambda)^2$ and with this precision is equal to
1.

For a special case $g_{xx}(\tau )=J^2\exp(-\lambda|\tau |);
g_{yy}(\tau )=0$ we reproduce the Kayanuma's result \cite{kay3}.

\section{Nondiagonal noise in spin-1  LZ theory.}

The Hamiltonian $H$ of a general multi-state LZ problem (see for
example \cite{brand} ) has the following matrix form
\begin{equation}
 H=Bt+A,
 \label{mlz}
\end{equation}
where $B$ is a diagonal matrix and matrices $A$ and $B$ are
independent on time. However, in a situation of a general position
only two levels cross. Several levels can intersect at the same
moment of time only due to a special symmetry. Such a symmetry is
systematically realized in the model of an arbitrary spin $S\neq
1/2$ placed into an external magnetic field that has a
time-dependent $z$-component vanishing at some moment of time
$t=0$ and a constant transverse component \cite{zeeman2}. The
corresponding Hamiltonian is:
\begin{equation}
H=\beta t S_z +g S_x
\label{arbh1}
\end{equation}
where $\beta$ and $g$ are constants. This  exactly solvable model
for a spin higher than $1/2$ was employed in the theory of Stark
effect \cite{Kaz}; some recent applications can be found in
\cite{{agu},{Suo}}.

In this section we generalize the result (\ref{probability}) to a
spin 1 system. We consider a following Hamiltonian for the spin 1
system in a random magnetic field:
\begin{equation}\label{hs1}
  H=(\beta t)S_z + \sum _{i=-1}^{+1} \eta_i S_i
\end{equation}
where $S_i$ are the spin projection operators for $S=1$. The
density matrix $\rho$ depends on 8 independent parameters. We
denote $\rho_+ = \rho_{1,1}-\rho_{0,0}$ and $\rho_{-} =
\rho_{-1,-1}-\rho _{0,0}$. The derivation of equations for $\rho
_+$ and $\rho _-$ can be done in the same spirit as it was done
for the spin $1/2$ case. Looking for a solution of the evolution
equation for density matrix in the form of perturbation series and
retaining after the averaging over the noise only the leading
terms in $1/\lambda$, we arrive at an integral-differential
equation. Additional care must be paid to the integrals of the
form $\int _{-\infty}^{t}dt_1 \int _{-\infty}^{t_1}dt_2
\exp(i\beta (t_1^2+t_2^2))g_{ij}(\lambda |t_1-t_2|)$ that did not
appear in the two level system, but appeared in the series for the
spin $S=1$. One can check that this integral is of the order
$1/\lambda$ and hence we disregard it and all terms that contain
it. After lengthy but straightforward calculations we find that in
the leading order in $1/\lambda$ the elements $\rho_+(t)$ and
$\rho_- (t)$ satisfy the following integral-differential equation:
\begin{equation}\label{v1}
\frac{d}{dt}
\begin{array}{l}
\left(
\begin{array}{l}
  \rho _+ (t) \\
  \rho _- (t)
\end{array}\right)= 
 -2\int _{-\infty}^{t}dt_1
\cos(\beta(t^2-t_1^2)/2)F(\lambda|t-t_1|)
 \left(\begin{array}{cc}
  1 & 1/2 \\
  1/2 & 1
\end{array} \right) \left( \begin{array}{c}
  \rho _+ (t_1)\\
  \rho _- (t_1)
\end{array} \right)
\end{array}
\end{equation}
The approximation of the fast noise allows to transform this
equation into the following differential one:
\begin{equation}\label{v1}
\frac{d}{dt} \left(
\begin{array}{l}
  \rho _+ (t) \\
  \rho _- (t)
\end{array}\right)=-2\hat{F}(\beta t)
\left(\begin{array}{cc}
  1 & 1/2 \\
  1/2 & 1
\end{array} \right)
\left( \begin{array}{c}
  \rho _+ (t_1)\\
  \rho _- (t_1)
\end{array} \right)
\end{equation}
Thus, the problem is reduced to a linear differential equation
with a constant matrix coefficient. The formal answer is:
\begin{equation}\label{r13}
\left( \begin{array}{l}
    \rho_{+}(t) \\
    \rho_{-}(t) \
  \end{array}\right) =\exp\left[-2\int_{-\infty}^t\hat{F}(\beta|\tau
  |)d\tau \left(\begin{array}{cc}
  1 & 1/2\\
  1/2 & 1
\end{array} \right)\right]
\left(
 \begin{array}{l}
  \rho _+ (-\infty)\\
  \rho _- (-\infty)
\end{array} \right)
\end{equation}
This formal solution can be transformed to a more explicit form
with the help of the matrix identity:
\begin{equation}\label{identity}
\exp\left[-\gamma \left(\begin{array}{cc}
  1 & 1/2\\
  1/2 & 1
\end{array} \right)\right]=
e^{-\gamma}
 \left( \begin{array}{cc}
  \cosh \gamma /2 & -\sinh\gamma /2\\
  -\sinh \gamma /2 & -\cosh\gamma /2
\end{array}\right)
\end{equation}
The transition probabilities are defined by this matrix acting on
vectors $\left( \begin{array}{l} 1 \\0\end{array} \right)$ if
initially the projection $+1$ was occupied , or
$\left(\begin{array}{l}0
\\1\end{array}\right)$ if the projection $-1$ was occupied. The
transition probabilities are:
\begin{equation}\label{tp}
  \begin{array}{l}
    P_{+\rightarrow 0} = P_{-\rightarrow 0} = \frac{1}{3} \left(1-e^{-3\gamma /2}\right) \\
    P_{+\rightarrow +}= P_{-\rightarrow -} = \frac{1}{3} + \frac{1}{6} e^{-3\gamma /2} +
    \frac{1}{2} e^{-\gamma /2} \\
    P_{+\rightarrow -} = P_{-\rightarrow +} = \frac{1}{3} + \frac{1}{6} e^{-3\gamma /2} -
    \frac{1}{2} e^{-\gamma /2}  \\
    P_{0\rightarrow 0} = \frac{1}{3} \left(1+e^{-3\gamma /2}\right) \\
    P_{0\rightarrow +} = P_{0\rightarrow -} =\frac{1}{3}\left(1-e^{-3\gamma /2}\right)
  \end{array}
\end{equation}
where $\gamma =\pi F(0)/\beta$.

We see that in the case of 3 levels the main results obtained in
the previous section for a 2-level system persist: the standard
white ($\gamma=\infty$) noise leads to equal population of all 3
levels, whereas the fast noise with a finite amplitude results in
non-trivial transition probabilities.

\section{Transition time for colored noise}

In section 2 we have shown that the typical time for establishing
the asymptotic of the transition probabilities is $\lambda/\beta
F(0)$. It is useful to look at the same problem from a different
point of view. Namely, we will analyze the behavior of transitions
driven by the standard $\delta$-like white noise, their typical
rates and times. If the action of the standard white noise is
limited by some finite time interval, it becomes physically
equivalent to the fast noise with a finite amplitude.
Simultaneously we will study directly the influence of the white
noise onto the LZ transitions.

Let us consider the problem with the very beginning. The
Hamiltonian is the generator of a random rotation:
\begin{equation}
H={\bf h(t)S} \label{hamiltonian}
 \end{equation}
where ${\bf h}(t)={\bf h}_0(t)+{\bf\eta}(t)$ and ${\bf
h_0}(t)=\beta t\hat{z}+\Gamma\hat{x}$.
Here we consider the isotropic white noise
$\langle\eta_i(t)\eta_k(t^{\prime})\rangle=\gamma\delta_{ik}
\delta(t-t^{\prime})$.\\
The density matrix as any Hermitian $S\times S$ matrix can be
represented as a sum:
\begin{equation}
{\bf\rho}=\rho^{(0)}I+\rho_i^{(1)}S_i+\rho_{ik}^{(2)}
\left[S_iS_k+S_kS_i-\frac{2}{3}S(S+1)\right]+...
\label{expansion}
\end{equation}
where the last term in equation (\ref{expansion}) contains $S$
operator factors. All irreducible tensors $\rho^{(j)};\,\,j=0...S$
evolve separately. The scalar $\rho^{(0)}=(1/N)Tr\rho$ is a
constant. The vector ${\bf\rho}^{(1)}$ obeys an obvious equation:
\begin{equation}
\dot{{\bf{\rho}}^{(1)}}=-{\bf h}\times\bf{\rho}^{(1)}
\label{vector}
\end{equation}
The modulus of the vector $\vec{\rho}^{(1)}$ is conserved.

Dynamic equation for the 2-nd order symmetric tensor $\rho_{ik}$
reads:
\begin{equation}
\dot{\rho}_{ik}=-\epsilon_{imn}h_m\rho_{nk}-\epsilon_{kmn}h_m\rho_{in}
\label{tensor}
\end{equation}
The extension for the rest of irreducible components is obvious.
In equation (\ref{vector}) for the vectorial part it is useful to
apply the interaction representation:
$\bf{\rho}^{(1)}(t)=U_0(t,t_0)\tilde{\bf{\rho}}^{(1)}(t)$, where
$U_0(t,t_0)$ is the evolution matrix for the field ${\bf
h}_0(t)=\hat{z}\beta t+\hat{x}\Gamma$. The reduced vector
$\tilde{\bf{\rho}}^{(1)}(t)$ obeys the following
equation:\begin{equation}
\frac{d}{dt}\tilde{\bf\rho}^{(1)}(t)=-\tilde{\bf\eta}
\times\tilde{\bf\rho}^{(1)}(t) \label{inter}
\end{equation}
Here $\tilde{\bf\eta}=U_0^{-1}({\bf\eta}\times U_0)$. In terms of
the Cartesian coordinates this equation reads:
\begin{equation}
\dot{\tilde{\rho}}_i=T_{ikl}\tilde{\rho}_k\eta_l
\label{components}
\end{equation}
where the matrix $T_{ikl}$ is:
\begin{equation}
T_{ikl}(t,t_0)=(U_0)_{mi}(U_0)_{nk}\epsilon_{mln}
\label{T}
\end{equation}
Initial values of ${\bf\rho}$ and $\tilde{\bf\rho}$ coincide since
$U_0(t_0,t_0)=I$. Equation (\ref{components}) can be solved by
power expansion over the noise:
\begin{equation}
\tilde{\rho}_i(t)=\rho_i(t_0)+\int_{t_0}^t T_{ikl}(t^{\prime},t_0)
\eta_l(t^{\prime})dt\rho_k(t_0)+...
\label{expansion}
\end{equation}
When averaging the expansion over the noise, all odd terms vanish.
To understand what happens with even terms consider first the
quadratic term:
$$\int_{t_0}^tdt_1\int_{t_0}^{t_1}dt_2T_{ik_1l_1}(t_1,t_0)T_{k_1,k,l_2}(t_2,t_0)
\langle\eta_{l_1}(t_1)\eta_{l_2}(t_2)\rangle\rho_k(t_0)$$

For the isotropic noise the contribution to the vector
$\tilde{\bf\rho}(t)$ up to the second order in ${\bf\eta}$ can be
represented as follows (we write equations for averages omitting
the angular brackets):
\begin{equation}
\tilde{\rho}_{2,i}(t)=\left(\delta_{ik}+\int_{t_0}^tdt_1M_{ik}(t_1)\right)\rho_k(t_0)
\label{quadratic}
\end{equation}
where
\begin{equation}
M_{ik}(t)=(\gamma /2) T_{iml}(t,t_0)T_{mkl}(t,t_0)=-\gamma\delta_{ik}
\label{M}
\end {equation}

Here we have used properties of the orthogonal matrices $U_0$:
$(U_0)_{mi}(U_0)_{mk}=\delta_{ik}$.

Next we proceed to the quartic term:
\begin{eqnarray}
\tilde{\rho}_{4,i}&=&\int_{t_0}^{t}dt_1\int_{t_0}^{t_1}dt_2\int_{t_0}^{t_2}dt_3
\int_{t_0}^{t_3}dt_4
T_{ipk}(t_1,t_0)T_{pql}(t_2,t_0)T_{qrm}(t_3,t_0)T_{rsn}
(t_4,t_0)\nonumber\\
&\times&\langle\eta_k(t_1)\eta_l(t_2)\eta_m(t_3)\eta_n(t_4)\rangle
\rho_{s}(t_0)
\label{quartic}
\end{eqnarray}
Quartic average of Gaussian field decays into quadratic averages:
\begin{eqnarray}
\langle\eta_k(t_1)\eta_l(t_2)\eta_m(t_3)\eta_n(t_4)\rangle=
\langle\eta_k(t_1)\eta_l(t_2)\rangle\langle\eta_m(t_3)\eta_n(t_4)\rangle\nonumber\\
 + \langle\eta_k(t_1)
\eta_m(t_3)\rangle\langle\eta_l(t_2)\eta_n(t_4)\rangle+\langle\eta_k(t_1)
\eta_n(t_4)\rangle\langle\eta_l(t_2)\eta_m(t_3)\rangle
\label{wick}
\end{eqnarray}
Only the first term contributes to the integral (\ref{quartic}).
Two others are zero at correct time ordering $t_1\geq t_2\geq
t_3\geq t_4$. In both cases considered earlier we obtain:
\begin{equation}
\tilde{\rho}_{4,i}=\int_{t_0}^{t}dt_1\int_{t_0}^{t_1}dt_2M_{ij}(t_1)M_{jk}(t_2)
\rho_k(t_0)
\label{M-square}
\end{equation}
The same result could be obtained from an effective
equation of motion for $\tilde{\rho}$:
\begin{equation}
\dot{\tilde{\rho}}_i=M_{ik}\tilde{\rho}_k
\label{effective}
\end{equation}
Thus, the operator ${\bf M}$ plays the role of effective
non-Hermitian Hamiltonian.

For the white isotropic noise the solution of the equation
(\ref{effective}) is:
\begin{equation}
\tilde{\rho}_i (t) = \tilde{\rho}_i (t_0) exp (-\gamma t)
\label{win}
\end{equation}
At $t \rightarrow \infty$ the formula (\ref{win}) always leads to
the occupation numbers $p=1/2$. As we know, this does not happen
for the colored noise with a finite amplitude. The reason is that
in the genuine LZ problem the solution strongly oscillates with
the frequency roughly $\omega(t) \sim \beta t$ long before and after
the level crossing point. This introduces a new energy scale that
must be compared with $\lambda$. For time in the range $|\beta
t|<\lambda$, the approximation of white noise is roughly valid
even for finite amplitude noise, but beyond this interval of time
the oscillations of the LZ solution become faster than the
correlation time of the noise, and the action of the noise is
suppressed by the oscillations.

In $Fig. 1$ we demonstrate a typical evolution of the probability for the system to stay in the same state as a
function of time. The evolution reminds diffusive motion that
slowly stops at large absolute values of time. The sharp change of
the absolute value of the amplitude near $t=0$ is due to the
constant transverse field.

\begin{figure}
\includegraphics{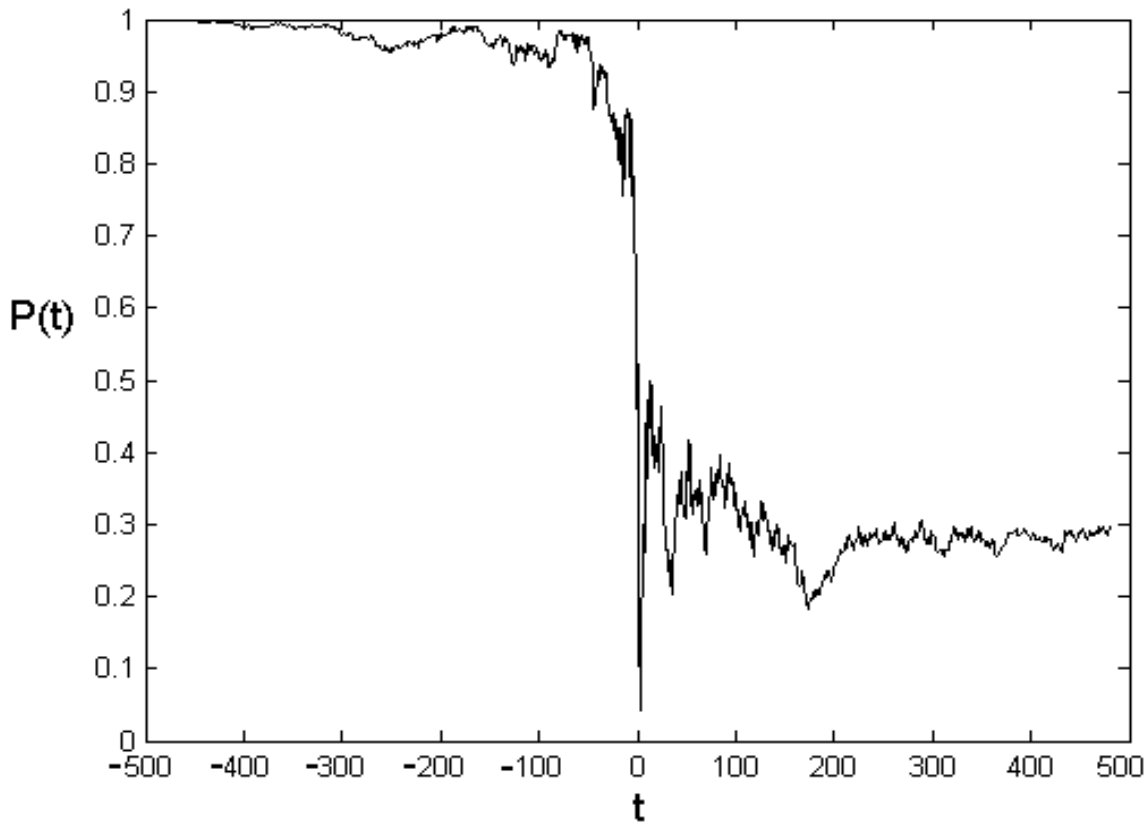}
\caption{Typical evolution of the  probability
$P(t)=|a_1(t)|^2$, where $a_1(t)$ is the amplitude of the first
diabatic state. The choice of parameters is as follows,
$g_{ij}= \delta_{ij} J^2 e^{-\lambda |t-t^{\prime}|}$, $\beta
=1$, $\lambda = 80$, $J^2 =0.18$, $\Gamma=0.7$.} \label{ff2}
\end{figure}
\noindent

To estimate roughly the transition probability one can apply the
standard white noise approximation in the time interval
$|t|<C\lambda/\beta$ and accept that at $|t|>C\lambda / \beta$ no
transitions due to the noise happen. The parameter $C$ is a
constant of the order of unity. Then we automatically
get the result $\rho (\infty) = \exp(-2C\lambda \gamma / \beta)\rho (-\infty)$.
 According to the definition, $\gamma \approx F(0)/ \lambda$ in
agreement with the calculations of the previous sections.
Summarizing, the transitions mediated by the fast noise proceed
during the large time interval of the order of $\lambda/\beta$.
This is the reason why its total effect remains finite, though it
is very small on a typical time scale of the usual LZ-transitions
mediated by constant field ($\tau \sim \Gamma/\beta$) \cite{Gef-time}.

\section{Landau-Zener transitions in a constant transverse field and in the colored noise.}
In previous sections we assumed that only the noise was
responsible for transitions, whereas the regular part of the
Hamiltonian operator was diagonal. In this section we incorporate
a regular non-diagonal operator (the transverse field) into the
Hamiltonian together with a noise. The
most general Hamiltonian for such a two-level system reads:
\begin{equation}\label{hig}
H=\beta t \sigma _z + g\sigma _x + \sum _i \eta _i (t) \sigma_i,\,\, i=x,y,z
\end{equation}
where $\eta _i$ are coordinates of the vectorial Gaussian noise with correlators 
given by (\ref{correlator}).

The second term in the Hamiltonian (\ref{hig}) can be considered
as a constant transverse magnetic field acting on a spin $S=1/2$.
The Hamiltonian (\ref{hig}) describes a spin (q-bit) weakly
interacting with the environment, for example with the nuclear
spin bath \cite{PS}. If this interaction is so small that the bath
relaxation is much faster than the inverse interaction energy, the
bath can be treated as a fast noisy magnetic field. The
measurements of the LZ transition probabilities can provide an
information about the strength of the coupling to the bath.
Another possible example is a molecular nanomagnet in fluctuating
dipole field.  The Hamiltonian (\ref{hig}) may be relevant to the
quantum shuttle problem where avoided level crossings occurred to
be important \cite{shuttle}. The fast noise in this example
corresponds to thermal fluctuations.

As it was discussed in the previous section, in the interval of
time $|\beta t|<<\lambda$ the spin experiences an equivalent white
noise with the amplitude $\gamma = F(0)/(2\lambda)$ vanishing at
$\lambda \rightarrow \infty$. The noise causes a slow decay of the
population difference $\rho = \rho_{11}-\rho_{22}$. The action of
the noise becomes noticeable only after long evolution time of the
order $|\lambda / \beta|$. On the contrary, the transitions due to
the constant transverse field proceed predominantly during a much shorter
interval of time $|t|<|g/\beta|$. During this interval of time the
role of noise is negligible comparing with the role of the
constant transverse field. Let us choose $t_0$ much less than
$|\lambda / \beta|$, but much larger than $g/\beta$ to make sure
that the transverse magnetic field is neglegible
 beyond the interval of time $(-t_0,t_0)$. The evolution from
$-\infty$ to $-t_0$ is influenced mainly by the noise. The result
of such an evolution is almost the same as that in the absence of
the constant transverse field with $t_0=0$. Since the effective
attenuation coefficient for $\rho$ is an even function of time
(see equation (\ref{diff})), the resulting exponent for $\rho
(t_0)$ is twice less than that for the problem of the section (2)
at $t=+\infty$ (equations (\ref{exp},\ref{infty}):
\begin{equation}
\rho(-t_0)=e^{-\pi F(0)/\beta}
 \label{r0p}
\end{equation}

In the interval of time $|t|<t_0$ the noise is neglegible and
transitions are completely due to the genuine Landau-Zener
mechanism, i.e. due to the constant transverse field $g$. Since
$t_0\gg g/\beta$, we can use the LZ formula to determine
$\rho(t_0)$. Let $a_1(t)$ and $a_2(t)$ be the amplitudes to find the
system at the first and the second diabatic states, respectively,
(to find the spin projection equals $1/2$ or $-1/2$) at the moment of
time $t$. Their values at $t=-t_0$ can be expressed in terms of
the population difference $\rho^*$ and two phase factors as
follows:

\begin{equation}
\begin{array}{l}
a_1(-t_0)=e^{i\phi_1}\sqrt{(1+\rho^*)/2} \\
a_2(-t_0)=e^{i\phi_2}\sqrt{(1-\rho^*)/2}
\end{array}
\label{aa}
\end{equation}
The phase diference $\phi_1-\phi_2$ is essentially random and
 independent on $\rho^*$ since,  the
solution of the Shr\"odinger equation strongly oscillates at $t<-t_0$ and
transitions are presumably due to the non-diagonal noise. The
amplitudes at time $t_0$ are related to those at $t=-t_0$ by a
linear relation $a_i(t_0)=S_{ij}(t_0,-t_0)a_j(-t_0)$, where $S$ is
the transition matrix in the noiseless LZ Hamiltonian from
$t=-t_0$ to $t=t_0$. Using this property and averaging over
the random phases $\phi_1$ and $\phi_2$ and over $\rho^*$ with
$<\rho^*>=\rho(-t_0)$, we arrive at a following result:

\begin{equation}
\rho (t_0)=<|a_1(t_0)|^2-|a_2(t_0)|^2>=e^{-\pi
F(0)/\beta}(2e^{-\pi |g|^2/\beta}-1) \label{ava}
\end{equation}
Here we employed the fact that our choise of time $t_0$ allows the approximations
$|S_{ij}(t_0,-t_0)| \approx |S_{ij}(\infty,-\infty)|$  and $\rho (-t_0) \approx \rho (0)$ 
The evolution from $t_0$ to $t=+\infty$ brings an additional
exponential factor equal to that in equation (\ref{r0p}):

 \begin{equation}
\rho(t \rightarrow \infty) = e^{-2\pi F(0)/\beta}(2e^{-\pi
|g|^2/\beta}-1)
 \label{rf}
 \end{equation}
 Correspondingly the probability to stay on the same diabatic level is
\begin{equation}\label{ff1}
  P_{1 \rightarrow 1}=\frac{1}{2}(1+e^{-2\pi F(0)/\beta}(2e^{-\pi |g|^2 / \beta}-1))
\end{equation}
In the limit of the noise only ($g=0$) and of zero noise
($F(0)=0$) the result (\ref{ff1}) transits into formula (\ref{probability})
or to the LZ formula, respectively.

The matching procedure used in this section is asymptotically
exact at $\lambda\rightarrow\infty$. To check how it works at
large but finite $\lambda$ we studied the LZ transitions subject
to a fast noise numerically.

\begin{figure}
\includegraphics{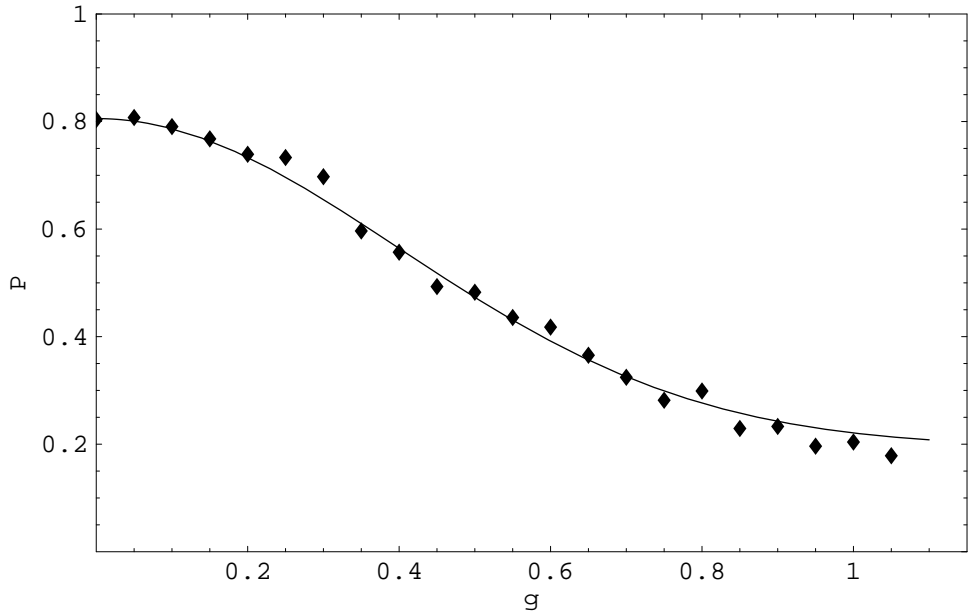}
\caption{The probability to stay on the same diabatic state in a
two level system at constant coupling to the noise as function of
transverse magnetic field. The choice of parameters is
$g_{ij}=\delta_{ix}\delta_{jx}J_x^2 e^{-\lambda |t-t^{\prime}|}$, $J_x=0.28$,
$\lambda = 125$, $\beta = 1$.} \label{ff2}
\end{figure}
\noindent We simulated the evolution generated by the Hamiltonian
(\ref{hig}). For simplification we put $\eta_y=\eta_z=0$ and
accept the following form for the correlator of the noise
$x$-component: $\langle\eta_x(t)\eta_x(t^{\prime})=J_x^2\exp
(-\lambda|t-t^{\prime}|)$. The time interval of the evolution was
chosen to be much larger than $\lambda/\beta$. In $Fig. 2$ we
depict the probability to stay in the same state after the level
crossing vs. the constant transverse field $g$ at a fixed coupling
to the noise $J_x=0.28$. Each discrete point represents the
averaging over 100 simulations with the same coupling constants.
The solid line is the graph of the theoretical formula
(\ref{ff1}). The deviations of the simulation results from the
analytical predictions do not exceed the accuracy of our
calculations. We conclude that equation (\ref{ff1}) describes well
the transition probability for the LZ system subject to a fast
noise.

\section{ Arbitrary spin in a strong diagonal random field}

We have demonstrated that the non-diagonal $\delta$-like white
noise leads to the equilibration of population between all states.
It is not correct for the white noise directed along the sweeping
field. Such a noise does not couple different states and its
action leads to the loss of coherence only. As it was shown
earlier, in the case of a 2-level system it results in a
Debye-Waller factor for $\rho$. We consider the case of a general
spin S placed into a regular field ${\bf h}_0=\hat{z}\beta t +
\hat{x} \Gamma$ and the random field directed along $z$-axis. Its
Hamiltonian reads:

\begin{equation}
H_{tot}=H+H_{noise}=\beta t S_z +\Gamma S_x +  \eta (t) S_z
\label{arbh2}
\end{equation}
We assume $<\eta (t) \eta (t ^{\prime})>=2\gamma \delta
(t-t^{\prime})$.

Following  \cite{SaiKay}, we expand the solution of equation
(\ref{arbh2}) in the power series over the noise amplitude and
average each term. The resulting series over powers of $\gamma$ is
a formal solution of a differential equation known as master
equation: \cite{SaiKay}

\begin{eqnarray}
\frac{\partial \rho (t)}{\partial t}=-i[H(t),\rho (t)]- \gamma
[S_z,[S_z,\rho (t)]]=\nonumber\\
-i[vtS_z+\Gamma S_x,
 \rho (t)]-\gamma (S_z^2 \rho (t)+\rho(t) S_z^2-2S_z \rho S_z)
\label{eq1}
\end{eqnarray}
It is convenient to introduce notations $\Gamma_{ij}=\Gamma
<i|S_x|j>$. Below we write down equation (\ref{eq1}) for diagonal
and non-diagonal elements of the density matrix separately:
\begin{equation}
\begin{array}{l}
\dot{\rho}_{ii}= -i(\Gamma_{i,i-1} (\rho _{i-1,i}-\rho _{i,i-1})
+\Gamma_{i,i+1} (\rho _{i+1,i}-\rho_{i,i+1}))\\
i=-S,-S+1, \ldots, S
\end{array}
\label{ri}
\end{equation}

\begin{equation}
\begin{array}{l}
\dot{\rho}_{mn}=-(i\beta t(m-n)+\gamma (n-m)^2) \rho _{mn}-\\
 -i (\Gamma _{m,m-1} \rho _{m-1,n} -\rho_{m,n+1} \Gamma _{n+1,n}+
\Gamma_{m,m+1} \rho_{m+1,n}-\rho _{m,n-1} \Gamma_{n-1,n}),\,\,\, m \ne n
\end{array}
\label{rii}
\end{equation}
It is possible to find an asymptotically exact solution of
equations (\ref{ri}), (\ref{rii}) in the limit of strong noise
$\gamma\gg\Gamma,\beta$. In this limiting case, non-diagonal
elements of the density matrix are $\sim\Gamma/\gamma$ times
smaller than diagonal ones. Indeed, let us disregard the dynamical
term $\dot{\rho}_{ij}$ in equations for the non-diagonal elements.
We will justify this approximation later. Then equation
(\ref{rii}) at $n=m\pm 1$ implies that the non-diagonal elements
$\rho _{n,n\pm 1}$ are suppressed comparing with diagonal matrix
element by the factor $\sim\Gamma/\gamma$. The matrix elements
$\rho_{n,n\pm 2}$ are suppressed by the same factor with respect
to $\rho_{n,n\pm 1}$ etc. The characteristic time interval
following from equation (\ref{ri})is $\Delta t\sim
|\rho_{nn}/(\Gamma\rho_{n,n\pm 1})\sim \gamma/\Gamma^2$. From this
estimate we find that the time derivative of the largest
non-diagonal matrix element
$|\dot{\rho}_{i,i\pm
1}|\sim(\Gamma^2/\gamma)|\rho_{i,i\pm 1}|\ll \gamma\rho_{i,i\pm 1}$
can be neglected. Retaining only main diagonal and two adjacent
non-diagonals in the matrix equations (\ref{ri}),(\ref{rii}), we express
the non-diagonal elements in terms of diagonal:

\begin{equation}
\begin{array}{l}
\rho _{i+1,i}=-\frac{i \Gamma _{i+1,i}}{i\beta t+\gamma}(\rho _{i,i}-\rho_{i+1,i+1}),\\
\rho _{i,i+1}=\frac{i \Gamma _{i,i+1}}{i\beta t-\gamma}(\rho _{i+1,i+1}-\rho_{i,i}),\\
\rho _{i,j}=0, \,\,  |i-j|>1
\end{array}
\label{diag}
\end{equation}
The problem is reduced to determining of the $2S+1$ diagonal
elements. Let introduce a vector ${\bf{c}}$ with the coordinates
$c_i=\rho _{ii}$. They are probabilities to find the spin in a
particular eigenstate of the operator $S_z$. Substitution of
(\ref{diag}) into (\ref{ri}) gives a differential equation for the
vector ${\bf{c}}(t)$
\begin{equation}
\dot{{\bf{c}}}(t)=(\frac{\Gamma^2}{i\beta
t+\gamma}-\frac{\Gamma^2}{i\beta t-\gamma})\hat{M} {\bf{c}}(t)
\label{de1}
\end{equation}
were the constant matrix $\hat{M}$ has following matrix elements:
\begin{equation}
\begin{array}{l}
 M_{ii}=-(<i+1|S_x|i>^2+<i|S_x|i-1>^2)=-\frac{1}{2}(S^2+S-i^2)\\
 M_{i,i+1}=M_{i+1,i}= <i+1|S_x|i>^2=\frac{1}{4}(S+i+1)(S-i)
\end{array}
\label{M}
\end{equation}
All other elements are zeros. Equation (\ref{de1}) can be easily
integrated
\begin{equation}
{\bf{c}}(t)=\exp (\int _{t_0}^{t} \frac{2 \gamma \Gamma^2 }
{(\beta t^{\prime})^2+\gamma ^2}dt ^{\prime} \hat{M}){\bf{c}}(t_0)
\label{ds}
\end{equation}
 To find transition probabilities we take limits of integral in
 equation (\ref{ds})  as $t_0=-\infty$ and $t=+\infty$:
\begin{equation}
{\bf{c}}(+\infty)=\exp (\frac{2 \pi \Gamma ^2}{\beta}
\hat{M}){\bf{c}}(-\infty) \label{cc}
\end{equation}
This result demonstrates that, for a general spin $S$ as well as
for a spin $1/2$, the transition probabilities do not depend on
the specific value of $\gamma$ provided that $\gamma$ is large.
Below we present explicitly the transition probabilities for some
values of spin. We denote $E_1=e^{-\Gamma ^2 \pi /\beta}$,
$E_2=e^{-3 \Gamma ^2 \pi /\beta}$ and $E_3=e^{-6 \pi \Gamma
^2/\beta}$. Then for $S=1/2$ the formula (\ref{cc}) reads
\begin{equation}
\begin{array}{l}
P_{1/2 \rightarrow 1/2}=P_{-1/2 \rightarrow -1/2}=\frac{1}{2}(1+E_1)   \\
P_{1/2 \rightarrow -1/2}=P_{-1/2 \rightarrow 1/2}=\frac{1}{2}(1-E_1)
\end{array}
\label{s12}
\end{equation}
For $S=1$
\begin{equation}
\begin{array}{l}
P_{1 \rightarrow 1}=P_{-1 \rightarrow -1}=\frac{1}{6}(2+E_2+3E_1)   \\
P_{1 \rightarrow 0}=P_{-1 \rightarrow 0}=P_{0 \rightarrow 1}=P_{0
\rightarrow -1}=\frac{1}{3}(1-E_2)   \\
P_{1 \rightarrow -1}=P_{-1 \rightarrow 1}=\frac{1}{6}(2+E_2-3E_1)   \\
P_{0 \rightarrow 0}=\frac{1}{3}(1+2E_2)
\end{array}
\label{s1}
\end{equation}
The results (\ref{s12}) and (\ref{s1}) coincide with already known
solutions for two- and three-level LZ models with strong
decoherence \cite{SaiKay}. A new result for a spin $S=3/2$ that
follows from (\ref{cc}) reads:
\begin{equation}
\begin{array}{l}
P_{3/2 \rightarrow 3/2}=P_{-3/2 \rightarrow -3/2}=
\frac{1}{4}+\frac{1}{20}E_3 +\frac{1}{4}E_2+\frac{9}{20}E_1\\
P_{3/2 \rightarrow 1/2}=P_{-3/2 \rightarrow -1/2}=
P_{1/2 \rightarrow 3/2}=P_{-1/2 \rightarrow -3/2}=
\frac{1}{4}-\frac{3}{20}E_3 -\frac{1}{4}E_2+\frac{3}{20}E_1\\
P_{3/2 \rightarrow -1/2}=P_{-3/2 \rightarrow 1/2}=
P_{-1/2 \rightarrow 3/2}=P_{1/2 \rightarrow -3/2}=
\frac{1}{4}+\frac{3}{20}E_3 -\frac{1}{4}E_2-\frac{3}{20}E_1\\
P_{3/2 \rightarrow -3/2}=P_{-3/2 \rightarrow 3/2}=
\frac{1}{4}-\frac{1}{20}E_3 +\frac{1}{4}E_2-\frac{9}{20}E_1\\
P_{1/2 \rightarrow 1/2}=P_{-1/2 \rightarrow -1/2}=
\frac{1}{4}+\frac{9}{20}E_3 +\frac{1}{4}E_2+\frac{1}{20}E_1\\
P_{1/2 \rightarrow -1/2}=P_{-1/2 \rightarrow 1/2}=
\frac{1}{4}-\frac{9}{20}E_3 +\frac{1}{4}E_2-\frac{1}{20}E_1\\
\end{array}
\label{s32}
\end{equation}

\begin{figure}
\includegraphics{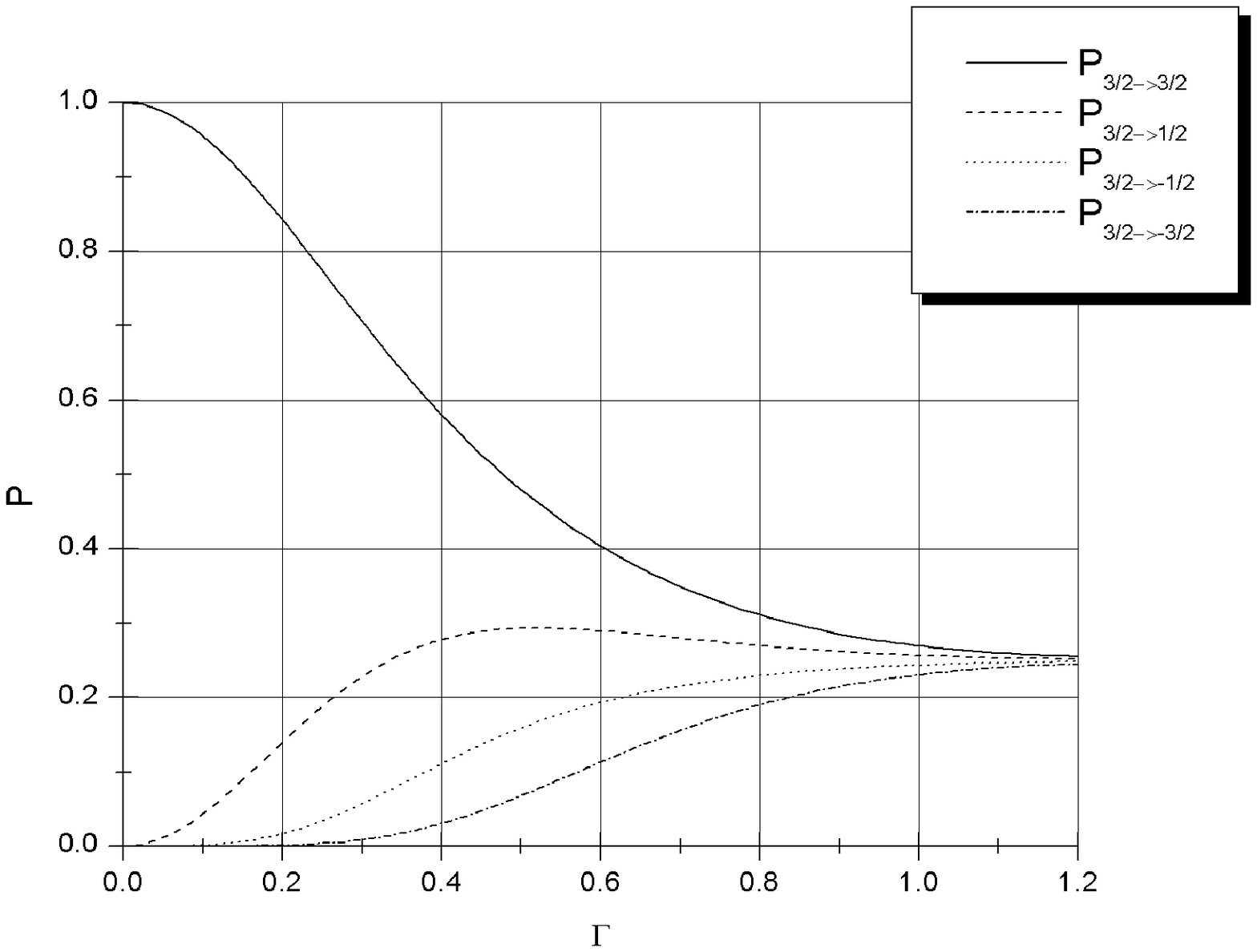}
\caption.{ Transition probabilities from the state with
$s_z=3/2$ to any other state as functions of transverse
magnetic field.}
\label{Fig1}
\end{figure}
\noindent

\begin{figure}
\includegraphics{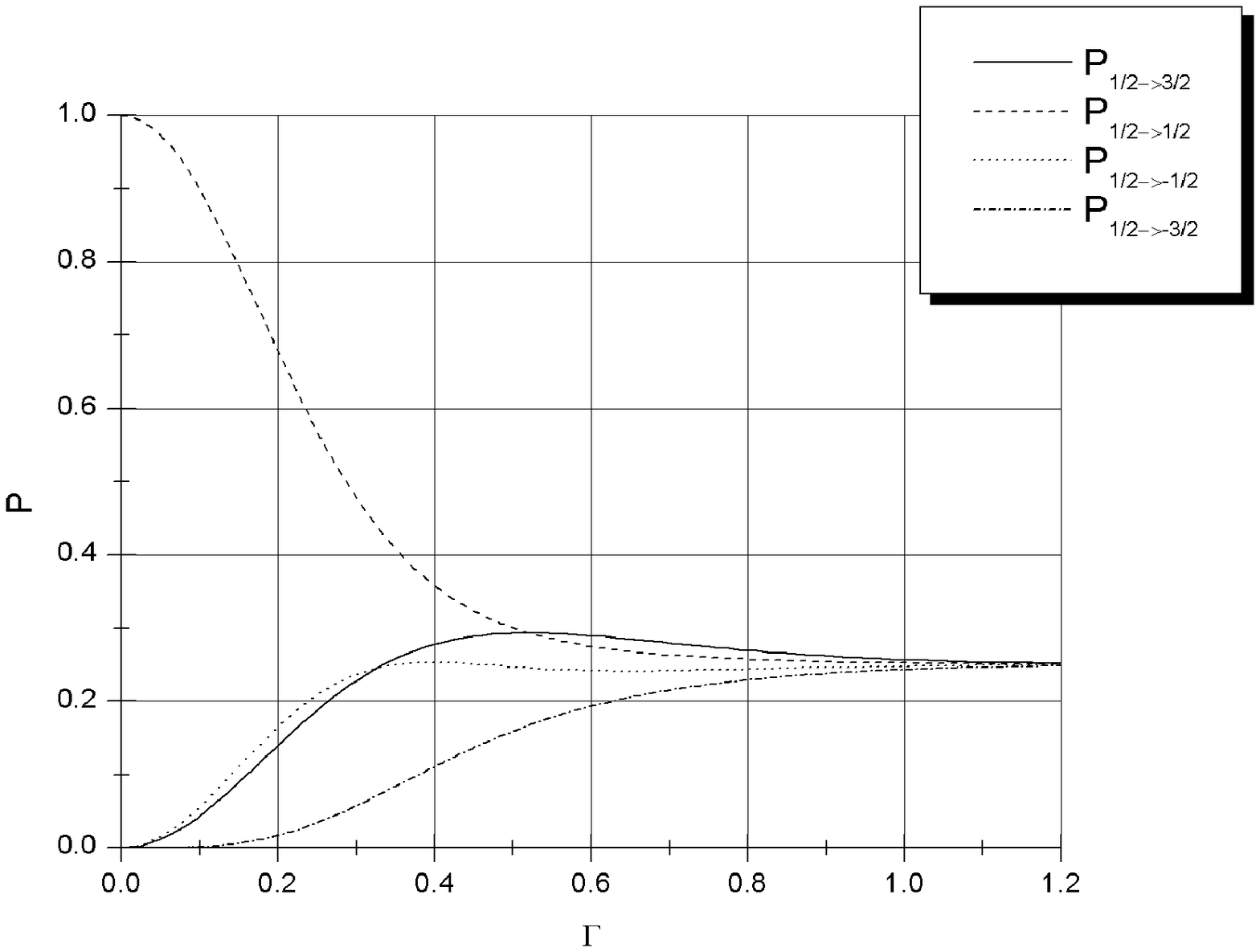}
\nopagebreak \caption{Transition probabilities from the state
 with $s_z=1/2$ to any other state as functions of transverse
 magnetic field.} \label{Fig2}
\end{figure}
\noindent

In $Fig3$ and $Fig4$  we show the dependences of transition
probabilities for $S=3/2$ on $\Gamma$. In adiabatic limit
$\Gamma^2/\beta >>1$ all states are equally populated after the
evolution.

\section{ Electron motion driven by electric field}.

Now we consider another multi-state Landau-Zener model that, in
the absence of the noise, was solved exactly \cite{PS}. Physically
the model describes the transport of a charged particle in a
regular linear chain driven by a time-dependent homogeneous
external field. Such a model is an idealization of atomic scale
molecular wires or linear arrays of quantum dots. An important
assumption in the model that makes it exactly solvable is that all
sites of the chain are identical and equidistant. External
electric field splits the energy levels at different sites of the
chain and suppresses the tunnelling between them. Hence the
transitions proceed in a narrow time intervals close to moments at
which the electric field becomes zero. The noise in such a system
arises due to thermal fluctuations chaotically changing the energy
of the electron. We suppose that there are no correlations of
noise at different sites.

Let us denote $|n \rangle$ a state located at the $n$-th site of the
chain. We assume that these states form a complete orthonormal set
(Wannier basis). In terms of this set the electron Hamiltonian
with linear dependence of an external field on time reads:

\begin{eqnarray}
\hat{H}= \sum\limits_{n} (g\mid n\rangle\langle n +
1\mid + c.c.) + n vt \mid n\rangle\langle n\mid  + \eta_n (t)
\mid n\rangle\langle n\mid
\label{h0}
\end{eqnarray}
where  $v$ and $g$ are constants and we assume that the noise
power is the same for all sites i.e. $<\eta_m \eta_n>=2 \gamma
\delta _{mn}$. The derivation of the master equation for this case
is similar to that of the previous section. We consider only the
limit of strong noise $\gamma>>g$. Then, as in previous example,
the non-diagonal elements of the density matrix $\rho _{i,j}$ with
$|i-j|>1$ can be neglected. Equations for diagonal matrix elements
of the density matrix are:
\begin{equation}
\dot{\rho_{nn}}=-ig(\rho_{n+1,n}+\rho_{n-1,n}-\rho_{n,n+1}-\rho_{n,n-1})
\label{dm1}
\end{equation}
Equations for non-diagonal elements after the averaging over the
random noise read:
\begin{equation}
\begin{array}{l}
\dot{\rho}_{n+1,n}=(-ivt-2\gamma )\rho _{n+1,n}-ig(\rho_{n,n}-\rho_{n+1,n+1})\\
\dot{\rho}_{n-1,n}=(ivt-2\gamma )\rho _{n-1,n}-ig(\rho_{n,n}-\rho_{n-1,n-1})
\end{array}
\label{nd2}
\end{equation}
Neglecting again time derivatives in these equations, we find:
\begin{equation}
\begin{array}{l}
\rho_{n+1,n}=\frac{-ig}{ivt+2\gamma}(\rho_{n,n}-\rho_{n+1,n+1})\\
\rho_{n-1,n}=\frac{ig}{ivt-2\gamma}(\rho_{n,n}-\rho_{n-1,n-1})
\end{array}
\label{nd3}
\end{equation}
Substituting (\ref{nd3}) into the equations (\ref{dm1}) for
diagonal elements, we obtain the evolution equations for diagonal
matrix elements of the density matrix $\rho _{n,n}$:
\begin{equation}
\dot{\rho}_{n,n}=g^2 \left( \frac{1}{ivt-2\gamma}-\frac{1}{ivt+2\gamma} \right)
(2\rho _{n,n}-\rho _{n+1,n} -\rho_{n-1,n})
\label{ch1}
\end{equation}
Without loss of generality we can assume that initially, at  $t =
- \infty$, the particle was located at the site number zero. It
can be treated as initial conditions for the master equation:
$\rho _{0,0}(t = -\infty)=1$ and all other elements of the density
matrix are zeros at $t = -\infty$. Then diagonal matrix elements
$\rho _{n,n}$ acquire the meaning of transition probabilities from
zeroth to the $n$-th site at a current time $t$. As in the
previous example we can find the solution for a chain of arbitrary
number of sites in the matrix form.

In the limit of infinite number of sites a compact solution can be
found by employing the Fourier-transformation $\rho
_{n,n}=\frac{1}{2\pi} \int _0^{2\pi} e^{in \phi} u(\phi,t)$. The
system of coupled differential equations (\ref{ch1}) is
diagonalized by this transformation. Corresponding differential
equation of the first order for the function $u(\phi,t)$ is
readily solved. Its solution with the initial condition
$u(\phi,-\infty =1)$ is
\begin{equation}
u(\phi,t)=\exp\left[-2\left(\frac{\pi}{2}+\arctan\frac{vt}{2\gamma}
\right)(1-\cos\phi)\right] \nonumber
\end{equation}
In the limit $t \rightarrow + \infty$ it approaches its limiting
value:
\begin{equation}
u(\phi,t \rightarrow + \infty)=e^{\frac{-4 \pi g^2}{v}(1-\cos \phi)}
\label{ch3}
\end{equation}
By the inverse Fourier-transformation we find the diagonal
elements of the density matrix:
\begin{equation}
\rho_{n,n} (t \rightarrow + \infty) = P_n=
e^{-\frac{4 \pi g^2}{v}}I_n(\frac{4 \pi g^2}{v})
\label{ch4}
\end{equation}
Here $P_n$ is the transition probability from the site with the
index $0$ to the cite with the index $n$ .

It is interesting to compare results of this calculation which
incorporates a strong noise with the transition probabilities
without noise. In the absence of the noise $(\gamma = 0)$ the
transition probabilities are \cite{PS}:

\begin{equation}
P_n^{(coh)}=|J_n(\sqrt{8 \pi}g)|^2
\label{ch5}
\end{equation}

\begin{figure}
\includegraphics{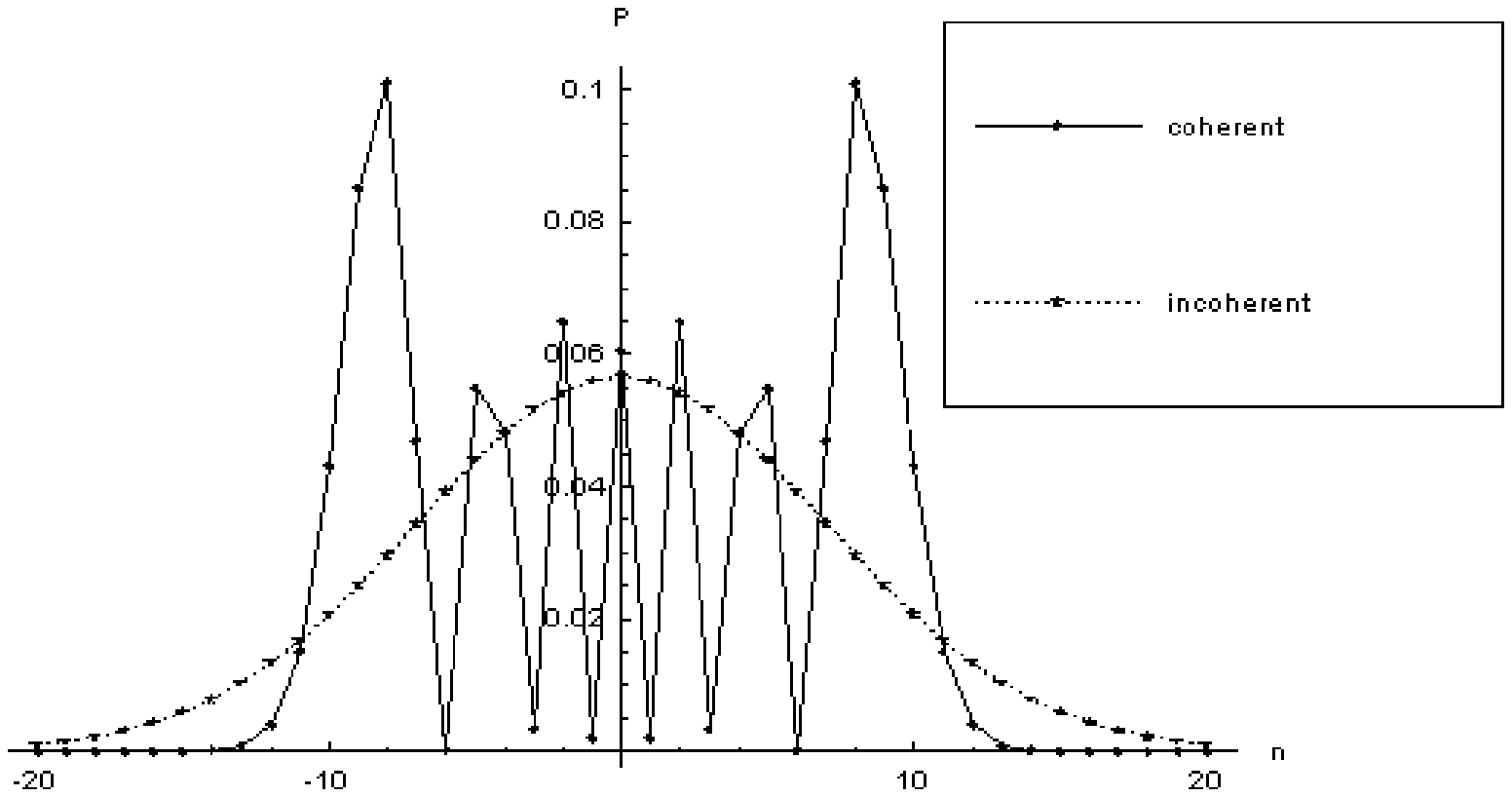}
\nopagebreak \caption{Comparison of transition probabilities in
 coherent (without noise) and incoherent (with strong noise) LZ
 models on the chain.}
\label{Fig2}
\end{figure}
\noindent

$Fig.5$ shows a typical behavior of the  transition probabilities for
both cases. The difference in the behavior is clearly pronounced.
In the absence of the noise the transition probabilities oscillate as
functions of $n$ and $g$ (see \cite{PS} for details). These oscillations
arise due to the interference among the amplitudes of different Feynman
paths leading from the initial to the final point. In the case of
a strong noise these oscillation are suppressed by the noise
imposed decoherence and the probability distribution is a smooth
bell-like curve. A simple parameter that is related to the effective
diffusion coefficient and can be measured experimentally is the
average square displacement of the particle during one sweep of
the external field. For a chain with a strong noise it is:
\begin{equation}
<n^2>=\sum \limits_{n=-\infty}^{+\infty}n^2 e^{\frac{-4 \pi g^2}{v}}
I_n(\frac{4 \pi g^2}{v})=\frac{4\pi g^2}{v}
\label{ch6}
\end{equation}
Despite a strong difference in the distribution functions, the
average square displacement (\ref{ch6}) coincides with that for
the coherent evolution without noise.

\section{Conclusion}

In conclusion, we have derived the formula for the transition
probabilities at the non-adiabatic crossing of two levels coupled
by a fast non-diagonal random field in the Landau-Zener
approximation. Depending on the strength of parameters it
interpolates between the Landau-Zener formula for noiseless system
and the Kayanuma's result for transitions mediated by the
transverse noise only. We have determined the time intervals
during which transitions are substantial and showed that time of
transitions mediated by the constant field is much shorter than
the time necessary for transition caused by the fast noise. Our
numerical simulations are in a good agreement with the analytical
formulas . We have discovered an important property of the
non-adiabatic tunnelling process: its probability depends not on
the "power" of the noise $\int_{-\infty}^{\infty}g(t)dt$ but
rather on the coupling to the noise $g(0)$. The former usually is
responsible for the decoherence rate in systems without time-
dependent fields. Measurements of the LZ transition probability
can provide the information about the value of this coupling. The
multi-level systems placed in regular time-dependent fields feel
subtle differences of the noise statistical properties.

We assumed that the diagonal noise does not dominate. In this
situation it does not play any role. However, if the component of
the diagonal noise is much stronger than non-diagonal ones, the
situation may change drastically. The problem of strong transverse field
compatible with noise inverse correlation time also remains open.

We have extended the formulas and methods employed for two-level
systems to a couple of multi-state LZ systems: an arbitrary spin
experiencing the time-dependent regular and random magnetic field
and a linear chain of sites in the external time-dependent
homogeneous electric field plus noise. The Landau-Zener
transitions for spins higher than $1/2$ were observed in a number
of systems \cite{{Kaz},{agu},{Suo}}. Therefore we believe that our
solutions can be checked experimentally.
\section{acknowledgements}

This work was supported by the NSF under the grant DMR0072115 and
DMR 0103455, by the TITF of Texas A$\&$M University and by the DOE
under the grant DE-FG03-96ER45598. We are grateful to M. Nishino for useful discussion.

\end{document}